\documentclass[conference]{IEEEtran}
\usepackage{cite}
\usepackage{siunitx} 
\usepackage{amsmath,amssymb,amsfonts}
\usepackage{graphicx}
\usepackage{listings}
\usepackage{textcomp}
\usepackage{xcolor}
\usepackage[most]{tcolorbox} 
\usepackage{algorithm}
\usepackage{algpseudocode}
\usepackage{tikz}
\usepackage{subcaption}
\usepackage{adjustbox}
\usepackage{caption}
\usepackage{setspace}
\usepackage{pgf-pie}  
\usepackage{makecell}
\usepackage{multirow}
\usepackage{url}
\usepackage{siunitx}
\usepackage{balance}
\usepackage{pgfplots}
\usepackage{tabularx} 
\usepackage{pifont}
\usepackage{pgfplots}
\usepackage{tabularx} 
\usepackage{graphicx}
\usepackage{booktabs}
\usepackage{enumitem}

\def\BibTeX{{\rm B\kern-.05em{\sc i\kern-.025em b}\kern-.08em
    T\kern-.1667em\lower.7ex\hbox{E}\kern-.125emX}}

\definecolor{codegreen}{rgb}{0,0.6,0}
\definecolor{codegray}{rgb}{0.5,0.5,0.5}
\definecolor{codepurple}{rgb}{0.58,0,0.82}
\definecolor{backcolour}{rgb}{0.95,0.95,0.92}
\definecolor{codered}{rgb}{0.91,0.58,0.53}

\definecolor{softblue}{RGB}{102,178,255}
\definecolor{softred}{RGB}{255,153,153}
\definecolor{softgreen}{RGB}{153,255,153}
\definecolor{softyellow}{RGB}{255,255,153}
\definecolor{softcyan}{RGB}{153,255,255}
\definecolor{softmagenta}{RGB}{255,153,255}
\definecolor{softorange}{RGB}{255,200,150}
\lstdefinestyle{mystyle}{
    backgroundcolor=\color{backcolour},   
    commentstyle=\color{black},
    keywordstyle=\color{black},
    numberstyle=\scriptsize\color{codegray},
    stringstyle=\color{black},
    basicstyle=\ttfamily\scriptsize, 
    breakatwhitespace=false,         
    breaklines=true,                 
    captionpos=b,                    
    keepspaces=true,                 
    numbers=left,                    
    numbersep=5pt,                  
    showspaces=false,                
    showstringspaces=false,
    showtabs=false,                  
    tabsize=2
}

\lstdefinestyle{mystyle_1}{
    backgroundcolor=\color{codered},   
    commentstyle=\color{codegreen},
    keywordstyle=\color{black},
    numberstyle=\scriptsize\color{codered},
    stringstyle=\color{codered},
    basicstyle=\ttfamily\scriptsize, 
    breakatwhitespace=false,         
    breaklines=true,                 
    captionpos=b,                    
    keepspaces=true,                 
    numbers=left,                    
    numbersep=5pt,                  
    showspaces=false,                
    showstringspaces=false,
    showtabs=false,                  
    tabsize=2
}
\newcommand{\tool}{\textsc{VistaFuzz}}
\newcommand{\bin}[1]{{\color{red} #1}}

\newcommand{\circled}[1]{\tikz[baseline=(char.base)]{
            \node[shape=circle,draw,inner sep=0.2pt] (char) {#1};}}

\begin{document}

\title{Harnessing LLMs for Document-Guided Fuzzing of OpenCV Library
\thanks{\bin{Identify applicable funding agency here. If none, delete this.}}
}


\author{
  Bin Duan$^{1}$,
  Tarek Mahmud$^{2}$,
  Meiru Che$^{3}$,
  Yan Yan$^{4}$,
  Naipeng Dong$^{1}$,
  Dan Dongseong Kim$^{1}$,
  Guowei Yang$^{1*}$ \\
  $^{1}$School of Electrical Engineering and Computer Science, The University of Queensland, Australia \\
  $^{2}$Department of Computer Science, Texas State University, USA \\
  $^{3}$College of Information and Communications Technology, Central Queensland University, Australia \\
  $^{4}$Department of Computer Science, University of Illinois Chicago, USA \\
  \{b.duan, n.dong, dan.kim, guowei.yang\}@uq.edu.au, \\
  {tarek\_mahmud@txstate.edu}, {m.che@cqu.edu.au}, {yyan55@uic.edu}\\
}

\maketitle

\begingroup
\renewcommand\thefootnote{*}
\renewcommand{\footnoterule}{}
\footnotetext{Corresponding author.}
\endgroup
\begin{abstract}
The combination of computer vision and artificial intelligence is fundamentally transforming a broad spectrum of industries by enabling machines to interpret and act upon visual data with high levels of accuracy.
As the biggest and by far the most popular open-source computer vision library, OpenCV library provides an extensive suite of programming functions supporting real-time computer vision. Bugs in the OpenCV library can affect the downstream computer vision applications, and it is critical to ensure the reliability of the OpenCV library.
This paper introduces \tool, a novel technique for harnessing large language models (LLMs) for document-guided fuzzing of the OpenCV library.
\tool\ utilizes LLMs to parse API documentation and obtain standardized API information. Based on this standardized information, \tool\ extracts constraints on individual input parameters and dependencies between these. Using these constraints and dependencies, \tool\ then generates new input values to systematically test each target API.
We evaluate the effectiveness of \tool\ in testing 330 APIs in the OpenCV library, and the results show that \tool\ detected 17 new bugs, where 10 bugs have been confirmed, and 5 of these have been fixed.
\end{abstract}


\section{Introduction}

Computer vision~\cite{szeliski2022computer, chai2021deep}, supported by libraries such as Open Source Computer Vision (OpenCV)~\cite{bradski2000OpenCV}, is changing the way machines interpret visual data. It plays an important role in areas such as facial recognition for security systems~\cite{prathaban2019vision}, gesture analysis~\cite{ismail2021hand}, and object detection~\cite{chandan2018real}. The high-level APIs of the OpenCV library provide an abstraction of the complex underlying computations and encapsulate sophisticated image processing algorithms~\cite{howse2020learning}. Developers can leverage 
these APIs without the need to delve into the intricacies of the supporting image processing algorithms. Underneath these 
APIs are the lower-level operations that perform a range of tasks, from basic image manipulation~\cite{minichino2015learning} to complex computer vision techniques like feature detection~\cite{noble2016comparison} and image stitching~\cite{ha2017evaluation}. 
Through these capabilities, in autonomous driving, OpenCV is used for lane detection in prototypes of self-driving vehicles~\cite{rossi2020real}; in the medical field, it is used for identification of medical imaging such as X-rays, magnetic resonance images~\cite{vu2023practical}, and CT scans~\cite{eswaran2024revolutionizing}; and it is also used for defect detection in automated assembly line products~\cite{baygin2017machine}, and providing navigational capabilities for drones in GPS-denied environments~\cite{couturier2019uav}.

Considering the need for AI applications and related computer vision models to function correctly and accurately, the reliability of computer vision systems is paramount, particularly when deployed in safety-critical applications~\cite{dong2021review}.
Yet, the complexity of the algorithms provided by the OpenCV library increases the risk of bugs that can be particularly stubborn and challenging to detect. 
Fuzzing~\cite{li2018fuzzing}, a powerful technique for finding bugs through random input generation, has been studied to test deep learning libraries~\cite{deng2023large,deng2024large} recently.  
Despite its promising results in testing deep learning APIs, it remains challenging to apply existing fuzzing techniques to test OpenCV library.

The essence of a fuzzing framework, API, or model lies in continuously generating valid inputs that meet requirements and are within certain boundaries~\cite{oehlert2005violating}. Thus, effective fuzzing requires an in-depth understanding of the constraints on the input, e.g., data types and sizes of the input parameters and input value ranges~\cite{eceiza2021fuzzing} to generate valid inputs.
We investigated the APIs in OpenCV and observed that \textit{\textbf{\circled{1}~many APIs in OpenCV library have dependencies between input parameters}}, which can greatly impact the validity of the generated test inputs. Precisely extracting constraints on individual input parameters and dependencies between different input parameters becomes crucial when generating inputs in an attempt to cover more possible behaviors of the API under test. Additionally, we notice that 
\textit{\textbf{\circled{2}~some APIs in OpenCV library lack descriptions of their input parameters.}}
Depending on how the OpenCV APIs are documented, they can be categorized into three types: (1) well-documented APIs, as shown in Listing~\ref{code:well-doc}, which has 399 such APIs; (2) poorly-documented APIs that have only API signatures but lack detailed descriptions, as shown in the Listing~\ref{code:poorly-doc}, which has 32 such APIs; and (3) undocumented APIs that have no documentation available at all, as shown in Listing~\ref{code:No-doc}, which has 248 such APIs. Notably, previous document-guided fuzzing methods~\cite{lv2020rtfm,xie2022docter} only focused on well-documented APIs, since they can not generate valid test cases that lack details of input parameters. 

Additionally, existing document-guided testing methods~\cite{liu2014automatic,zhou2018automatic,blasi2018translating,lv2020rtfm,xie2022docter} have not been explored for OpenCV, as they do not account for its strict parameter dependencies. Meanwhile, LLM-based testing approaches~\cite{deng2023large,deng2024large,meng2024large,xia2024fuzz4all} lack sufficient understanding of OpenCV, making it difficult to generate valid test cases, limiting their effectiveness in fuzzing.

\lstset{style=mystyle}
\label{code:well-doc}
\begin{lstlisting}[language=Python, caption= Well-documented API, label=code:well-doc]
cv2.getRotationMatrix2D.__doc__:
'getRotationMatrix2D(center, angle, scale) -> retval
. @brief Calculates an affine matrix of 2D rotation.
. ...
. @param center Center of the rotation in the source image.
. @param angle Rotation angle in degrees. Positive values mean counter-clockwise rotation.
. @param scale Isotropic scale factor.
. ...'
\end{lstlisting}

\lstset{style=mystyle}
\label{code:poorly-doc}
\begin{lstlisting}[language=Python, caption= Poorly-documented API, label=code:poorly-doc]
cv2.calcBackProject.__doc__:
'calcBackProject(images, channels, hist, ranges, scale[, dst]) -> dst'
\end{lstlisting}

\lstset{style=mystyle}
\label{code:No-doc}
\begin{lstlisting}[language=Python, caption= Undocumented API, label=code:No-doc]
cv2.aruco.__doc__:
'No documentation available'
\end{lstlisting}

To address these issues, we propose \tool, the first fuzzing technique to test OpenCV library. \tool\ leverages LLM to parse API documentation into standardized API information, and learns information of input parameters from well-documented APIs, and generates similar standardized information for poorly-documented APIs.
Based on standardized API information, we extract constraints on individual input parameters and dependencies between these parameters, we generate the input values (aka. test cases) that satisfy these constraints and dependencies for fuzzing. If an unexpected output is detected during the testing process, the potentially problematic API and its corresponding input test case are reported for further investigation. 
Thus, \tool\ can generate test cases that meet the requirements specified in the API documentations for fuzzing, thereby effectively testing the OpenCV library.

To evaluate the effectiveness of \tool, we tested 330 APIs in OpenCV library using \tool. 
As a result, \tool\ detected a total of 17 new bugs, where 10 have been confirmed and 5 have been fixed.

In summary, this paper makes the following contributions: 
\begin{itemize}[leftmargin=*]
\item We introduce \tool, a novel document-guided fuzzing approach for testing OpenCV library. To facilitate effective fuzzing, \tool\ harnesses LLMs to parse and learn API documentation to generate standardized API information, from which it extracts constraints on each input parameter and dependencies between input parameters and thereby generates valid input values. 
To the best of our knowledge, this is the first work on automated testing of OpenCV library.

\item We develop a prototype \tool\ using GPT-4.
Our tool, along with standardized API is  publicly available to facilitate the replication and more extensive evaluation of 
\tool\footnote{\tool, {https://github.com/beanduan22/VistaFuzz}}.

\item The evaluation of \tool\ on testing 330 OpenCV APIs shows that \tool\ detected a total of 17 bugs in OpenCV (v4.9.0). In particular, 10 new bugs have been confirmed, and 5 of them have been fixed in the latest version (v4.11.0).

\end{itemize}

\section{Background}
\subsection{OpenCV Library}

Computer vision~\cite{voulodimos2018deep, szeliski2022computer, xu2021computer} has rapidly become one of the fastest-growing branches of computing. Various libraries~\cite{riba2020kornia, niitani2017chainercv, cazorla2015javavis, dehghani2022scenic, handa2016gvnn} offer powerful image-processing algorithms to aid application development. The most popular for general image processing is the OpenCV library~\cite{bradski2000OpenCV}. OpenCV provides a robust platform for real-time image processing, feature detection, and object recognition~\cite{golovnin2021benchmarking, hasan2021face, giri2022emotion}, enabling complex visual comprehension. It is a primary tool in machine learning and AI studies, facilitating research in object detection~\cite{sharma2021object}, facial recognition~\cite{khan2020real}, and automated visual inspection~\cite{affolder2022automated}.
Given the critical demand for precision and reliability in these fields, ensuring the robustness of these APIs is essential.
Previously, there were many works on testing deep learning applications, such as deep learning API~\cite{xie2022docter, deng2023large} and deep learning compiler~\cite{xiao2022metamorphic, liu2023nnsmith}, but none of them involved testing OpenCV. 

This paper is the first work to test the OpenCV APIs. Since most popular deep learning libraries are written in Python, we focus on OpenCV-python, the interface provided by the OpenCV library for Python. Furthermore, testing the OpenCV-python API allows for the evaluation of the Python interface and a thorough testing of the underlying C++ implementation called by OpenCV-python.

\subsection{Large Language Models}

Since the introduction of the Transformer~\cite{kitaev2020reformer} architecture, LLMs have revolutionized language understanding and generation~\cite{zhu2023minigpt}, excelling in tasks like natural language understanding~\cite{min2023recent}, pattern recognition~\cite{guo2023images}, and transfer learning~\cite{xiao2023offsite}. Trained on vast textual data, LLMs adeptly identify and comprehend linguistic patterns, common expressions, and technical terminologies, enabling them to extract information, understand context, interpret complex language structures~\cite{chang2023survey}, and adapt to domain-specific applications such as code generation~\cite{chen2021evaluating} and automated reasoning~\cite{bubeck2023sparks}.

In this paper, we utilize the capabilities of GPT-4 to pre-parse and learn the API information from the OpenCV official documentation and standardized input and output parameters in these APIs. Additionally, based on the ability to understand linguistic patterns and perform transfer learning, GPT-4 can infer and supplement missing details for APIs that lack comprehensive documentation. This approach allows us to effectively obtain intermediate representations of different APIs from the official documentation, thereby creating consistent and standardized API information.

\begin{figure*}[t!]
  \centering
  \includegraphics[width=0.8\textwidth]{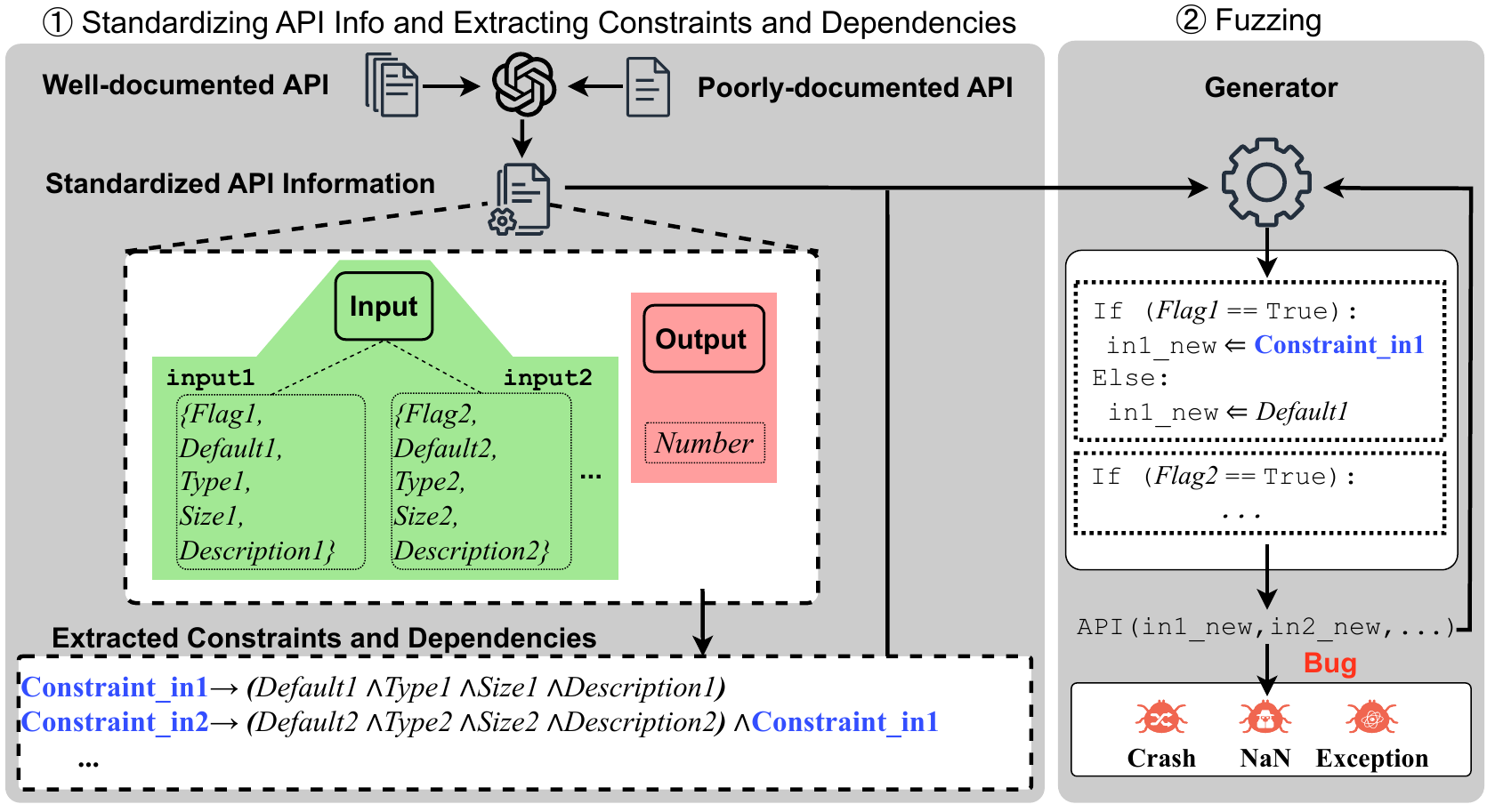}
  \caption{Overview of \tool}
  \label{fig:overview}
\end{figure*}

\section{Approach}

Figure~\ref{fig:overview} shows the overview of our approach \tool. It first generates standardized API information using GPT-4 and extracts constraints and dependencies from these. For well-documented APIs, GPT-4 can directly parse the information; for poorly-documented APIs, GPT-4 standardizes them by learning the information of parameters from well-documented APIs. Next, we start to extract constraints and dependencies for generating input test cases from standardized API information, which are then leveraged to generate test cases and perform fuzzing on target APIs in OpenCV library to detect three common types of bugs in APIs, i.e., crash, nan, and unexpected exception.

\begin{figure}[t!]
  \centering
  \includegraphics[width=0.85\linewidth]{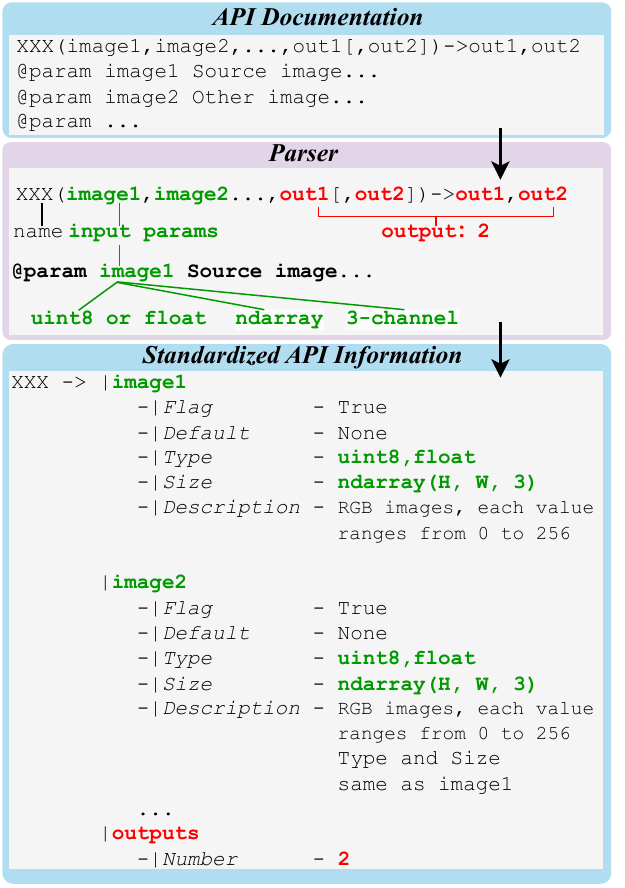}
  \caption{The example of parser}
  \label{fig:parser}
\end{figure}

\subsection{Standardizing API Information}
\label{sec:3.1}

Due to the complexity of the OpenCV APIs, e.g., the input parameters of an API have their individual constraints and also have dependency on each other, existing automated testing tools face challenges in generating valid test inputs to test these APIs in the OpenCV library. To address these challenges, we have implemented a solution based on the official documentation and GPT-4. 
Initially, GPT-4 was utilized to parse well-documented APIs, which contributed to understanding the functions and features of the parameters, laying the groundwork for extracting constraints and dependencies among them. As our research evolved, we found that GPT-4 not only generates standardized API information but also learns parameter information from these APIs. This capability allowed us to infer information of poorly-documented APIs effectively, generating standardized information that meets our expectations. This approach facilitates the construction of comprehensive, standardized API information, simplifying automated testing and broadening the coverage of APIs, thereby minimizing manual intervention and enhancing the reliability of the testing process.

We begin with well-documented APIs, as illustrated in Figure~\ref{fig:parser}'s \textbf{\small\textit{API Documentation}}, which provides function signatures and parameter descriptions. This documentation is processed by GPT-4 according to predefined rules for extracting API names, input parameters, and output parameters. 
As shown in Figure~\ref{fig:parser}'s \textbf{\small\textit{Parser}}, the function signature's initial string is recognized as the API name. Parameters enclosed in brackets and appearing after an arrow are classified as output parameters, which are not further analyzed in detail. The remaining parameters are treated as input parameters. 
For these input parameters, as shown in Figure~\ref{fig:parser}, the parameter \texttt{\small{'image1'}}, typically a three-channel array, is defined to be of type \texttt{\small{uint8}} or \texttt{\small{float}}, which corresponds to an RGB image. Following these parsing rules, GPT-4 generates standardized API information, as illustrated in Figure~\ref{fig:parser}'s \textbf{\small\textit{Standardized API Information}}. This standardized format includes API names, input parameter details, descriptions, and the number of output parameters.
Each input parameter is characterized by five attributes:
\begin{itemize}
    \item \textit{Flag}, indicates whether the parameter is modifiable.
    \item \textit{Default}, specifies whether it has a predefined value. 
    \item \textit{Type}, defines the data type of the parameter.
    \item \textit{Size}, defines the data structure and dimensions to ensure compatibility with the API requirements. 
    \item \textit{Description}, provides details on the acceptable value range and whether the parameter is influenced by other parameters.
\end{itemize}
This standardized API information enhances consistency and facilitates automated test case generation.

However, as mentioned earlier, not all APIs are well-documented. Incomplete documentation may indicate unreliable functionality and a higher likelihood of containing bugs. Figure~\ref{fig:parser} illustrates our approach to parsing well-documented APIs. 
For instance, APIs like \texttt{\small{cv2.getRotationMatrix2D}} (Listing~\ref{code:well-doc}) include comprehensive documentation elements such as \texttt{\small{@brief}}, which explains the computational logic, and \texttt{\small{@param}}, which provides detailed descriptions of each parameter. GPT-4 can parse this information directly from the documentation.
In contrast, poorly-documented APIs, such as \texttt{\small{cv2.calcBackProject}} (Listing~\ref{code:poorly-doc}), lack these detailed descriptions. To address this, we leverage GPT-4’s ability to infer missing details by referencing standardized information from well-documented APIs. 
For example, if the poorly-documented API \texttt{\small{cv2.calcBackProject}} contains a parameter named \texttt{\small{scale}}, GPT-4 identifies similar parameters from well-documented APIs, such as \texttt{\small{cv2.getRotationMatrix2D}} (Listing~\ref{code:well-doc}), and infers its likely meaning based on context. This approach ensures consistency by filling in missing information while maintaining a standardized format across APIs.

By systematically generating standardized API information, our method reduces manual effort, enhances the reliability of automated testing, and expands API coverage, which can detect bugs hidden in poorly-documentation APIs.

The specific prompts consist of the following:
\begin{itemize}
    \item \textbf{Input:} Provide the raw API documentation, including API signature and detailed information of each parameter (if available).
    \item \textbf{Task:} Parse the raw API documentation to generate a standardized API information by performing the following steps:
    \begin{enumerate}
        \item Identify the API name from the function signature.
        \item Classify parameters as input or output based on syntactic markers (e.g., arrows or brackets in the signature).
        \item Generate a standardized API format containing:
        \begin{itemize}
            \item API name.
            \item Input parameters with \textit{Flag}, \textit{Default}, \textit{Type}, \textit{Size}, and \textit{Description}.
            \item Number of output parameters.
        \end{itemize}
         \item For well-documented APIs, the standardized information is generated from the given documentation. 
         \item For poorly-documented APIs, missing details are inferred using patterns from well-documented APIs.
    \end{enumerate}
    \item \textbf{Output:} Standardized API information that adheres to the above requirements, ensuring consistency across APIs.
\end{itemize}

Additionally, we provided specific examples to help GPT-4 understand our requirements and the expected output format. After a few-shot learning process, GPT-4 was able to accurately generate the standardized API information..

\subsection{Constraints and Dependencies Extraction}
\label{subsec:constraints}

\begin{figure*}[h]
  \centering
  \includegraphics[width=\textwidth]{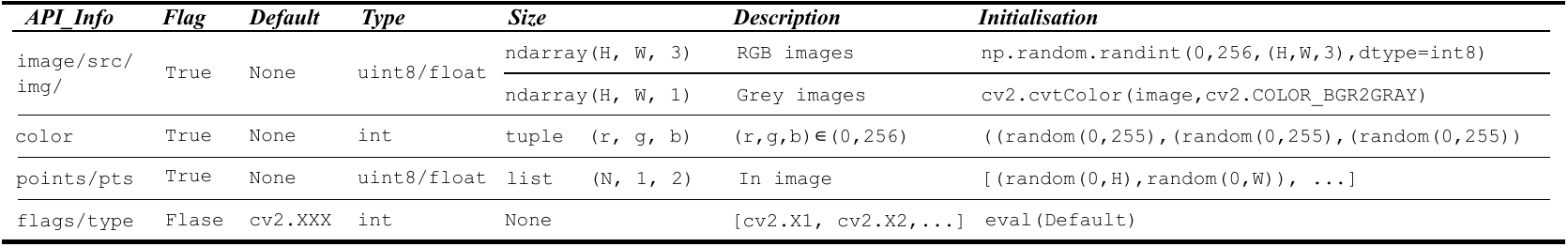}
  \caption{Example of Standardized API Information and Initialisation}
  \label{fig:example_2}
\end{figure*}

After obtaining standardized API information, \tool\ further extracts constraints and dependencies to generate test cases.

First, we extract constraints for each input parameter from the standardized API information, which includes five key aspects: \textit{Flag}, \textit{Default}, \textit{Type}, \textit{Size}, and \textit{Description}, as shown in Figure~\ref{fig:overview}.

We begin by checking \textit{Flag}. If \textit{Flag} is False, we assign it the value from \textit{Default}. If it is True, we use \textit{Type} to determine the data type of the parameter. For example, consider the input parameter \texttt{\small{'image1'}} in Figure~\ref{fig:parser}'s \textbf{\small\textit{Standardized\ API\ Information}}, which supports elements of type uint8 and float. Accordingly, we set its type based on this information. Next, we use \textit{Size} to define the data structure and dimensions of the input parameters. In this example, \texttt{\small{'image1'}} is a three-dimensional array with three channels and customizable height and width.

Subsequently, \textit{Description} further refines constraints. If no additional details are provided, we default to a standard three-channel RGB image of size $H\times W\times 3$. If the documentation specifies grayscale or another format, we adjust accordingly to a single or dual-channel image.

For dependent parameters, we account for both their intrinsic constraints and dependencies on other parameters, particularly in \textit{Type} and \textit{Size}. For instance, in Figure~\ref{fig:parser}, if \texttt{\small{'image2'}} references \texttt{\small{'image1'}} in its \textit{Description}, we enforce dependencies to maintain correctly generation. 

As shown in Figure~\ref{fig:example_2}, parameters like  \texttt{\small{'color'}} are randomly initialized within valid ranges of 0 to 256, while \texttt{\small{'points/pts'}} are constrained within image dimensions. If \textit{Flag} is True, the parameter is modifiable following our generation strategies; otherwise, it is selected strictly within its its predefined constraints in \textit{Description}, as exemplified by \texttt{\small{'flags/type'}} in Figure~\ref{fig:example_2}. This process systematically ensures valid parameter generation, reducing errors caused by non-standard inputs.

\subsection{Fuzzing on OpenCV APIs}
\label{sec:fuzzing}

In this section, we explore how fuzzing can be applied to test OpenCV APIs. Our approach generates test cases using standardized API information and extracted constraints and dependencies, ensuring that input values cover diverse scenarios, including edge cases and extreme conditions.

Fuzzing the OpenCV APIs, as part of broader API fuzzing strategies, typically involves generating large-scale test cases and monitoring API responses. This process not only verifies the functionality and performance of the APIs but also ensures reliability by testing whether APIs correctly handle unexpected or extreme inputs. During these tests, parameters are fed into the API, and the response data is scrutinized to confirm that the API's behavior aligns with predefined performance benchmarks. 
Here, our focus is on functional testing to detect bugs, anomalous behaviors, or potential vulnerabilities that may not be evident under normal operating conditions but could emerge in maliciously engineered environments.

\subsubsection{Generation Strategies}
\label{sec:gs}

Before generating test cases, we first determine the \textit{{Flag}} status of each input parameter in the standardized API information. If \textit{{Flag}} is \textit{False}, the parameter is excluded from generation strategies, indicating that it has constrained values specified in \textit{{Description}}. In this case, values are randomly selected within predefined constraints. If \textit{{Flag}} is \textit{True}, we proceed with test case generation strategies.
Our test case generation follows strategies commonly used in software testing~\cite{wei2022free}, covering three key aspects: \textit{Type}, \textit{Size}, and \textit{Value}.

\textbf{\textit{Type:}}
This step assigns valid data types to input parameters while ensuring dependencies remain consistent. For parameters that support multiple types (e.g., uint8, float32, float64), we test different valid types. If a parameter only allows one type, it remains unchanged. We also introduce invalid types (e.g., passing a string where a number is expected) to check whether the API correctly handles type mismatches. Ensuring type consistency is important because some parameters depend on others—for example, if one parameter is an image array, related parameters should match its type. Through these methods, our goal is to ensure that the API maintains reliable functionality, handling both valid and invalid inputs without crashing or producing undefined behavior.

\textbf{\textit{Size:}}
This step determines the size of input parameters while considering their constraints and dependencies. For different data structures (arrays, tuples, lists), we can adjust their dimension to make the generated input parameters more diverse. For example, random values \textit{H} and \textit{W} are assigned as height and width. RGB images use the size (\textit{H}, \textit{W}, 3), while grayscale images are converted into single-channel images of size (\textit{H}, \textit{W}, 1) using \texttt{\small{cv2.cvtColor}}. List-type parameters, such as the points, require only \textit{H}, setting size as (\textit{H}, 1, 2). For parameters with dependencies, sizes are assigned based on earlier values. Fixed-size parameters (e.g., "color") remain unchanged.

Furthermore, to ensure reliability of OpenCV APIs, we introduce extreme condition testing by introducing exceptionally large, small, or malformed data inputs. This evaluates their resilience and ability to handle errors, which is vital for reliable and secure operation in computer vision applications under adversarial or unusual conditions.

\textbf{\textit{Value:}} Value generation involves manipulating the actual numerical values of input data within the constraints of input type and size to assess the reliability of the API against diverse data types and adversarial conditions. We employ the following methods of value generation to simulate real-world scenarios and test the system's resilience:

\begin{itemize}[leftmargin=*]
\item{Adding Noise:} By introducing various types of noise into the test case (e.g. Gaussian noise), we can evaluate the API's capability to handle imperfect input data containing noise. This method helps reveal how the API performs in real-world applications facing data quality issues.
\item{Random Masking:} We opt to partially or completely mask input parameters with fixed integer pixel values. This approach simulates scenarios where images might be obstructed, testing the API's effectiveness in processing incomplete image information.
\item{Division:} Dividing image pixel values by an integer tests the API's ability to handle scaling or intensity changes. This operation evaluates the API's sensitivity to changes in image brightness or contrast.
\end{itemize}

These strategies help us assess the API’s stability and accuracy under various input conditions, particularly when handling dynamic changes in image quality. By identifying potential bugs, we ensure the API API maintains robust performance across a broad spectrum of real-world applications, thereby enhancing the reliability of the system.

\subsubsection{Fuzzing Automation}
\label{algorithm}

\begin{algorithm}[t!]
\footnotesize
\caption{Fuzzing}
\label{algo1}
\begin{algorithmic}[1] 
\Procedure{\tool}{$Standardized\_Info$}
\State $Input: Initialized\_Input, Standardized\_Info$  
\State $Output: BugInput, BugAPI$  
    \State $Pars, cv2.API \Leftarrow Standardized\_Info$
    \State $New\_Args = Initialized\_Input$
    \While{$condition$}
    \For{$Par$ \textbf{in} $Pars$}
        \State $ParInfo = Standardized\_Info['Par']$
        \State $Now\_Arg = New\_Args['Par']$
        \If{$ParInfo['Flag']$ == $False$}
            \State$ New\_Arg = Random(ParInfo['Description'])$
        \Else
            \If{$ParInfo['Type']$ \textbf{is} $None$}
                \State $N\_Type = Now\_Arg.type() $
            \ElsIf{$Dependencies$ }
                \State $N\_Type = Dependencies['Type']$
            \Else
                \State $N\_Type = TYPE(ParInfo['Type'])$
            \EndIf
            \If{$ParInfo['Size']$ \textbf{is} $None$}
                \State $N\_Size = Now\_Arg.size() $
            \ElsIf{$Dependencies$}
                \State $N\_Size = Dependencies['Size']$
            \Else
                \State $N\_Size = SIZE(ParInfo['Size'])$
            \EndIf
            \State $Strategy = Random(Value\_Strategies)$
        \EndIf
        \State $ New\_Arg = Gen(Now\_Arg, N\_Type, N\_Size, Strategy)$
        \State $ New\_Args['Par'].update(New\_Arg)$
    \EndFor
    \State $Outputs = cv2.API(New\_Args)$
    \If{$Bug$}
        \State $Record (New\_Args, cv2.API)$
    \EndIf
\EndWhile
\EndProcedure
\end{algorithmic}
\end{algorithm}

Fuzzing is conducted within a framework of specific fault tolerance and iteration limits, continuously generating test cases, and recording crashes, NaN, or tolerance violations that occur. We focus on test cases that lead to unexpected outcomes and document any anomalies for subsequent analysis. If all test cases successfully complete the stipulated number of attempts, the tested API is deemed correct. Each of these strategies can reveal different categories of defects: Type may uncover issues with type checking or how the API fails upon encountering unexpected data types. Size could expose bugs related to data handling, such as how the API manages memory and processes data of unexpected lengths. Value is crucial for understanding the API's logic validation and whether it can correctly handle a wide range of input values. Implementing these strategies requires an understanding of the API's schema, the expected input range, types, and behaviors. It is about creating tests that push the boundaries of these expectations to ensure that the API remains robust under a variety of inputs that could occur in real-world scenarios.

Algorithm~\ref{algo1} outlines the key steps in \tool. We first retrieve the standardized API information and input arguments (lines 1-2). We next initialize the test case, and extract the target API along with its parameter list \textit{Pars} (line 4), then obtain the initial input arguments\textit{New\_Args} (line 5). The fuzzing process runs with a predefined limit on test case generation, terminating once this limit is reached (line 6). 

Initially, we iterate through each parameter in the current API input parameter list (line 7). From the standardized API information, we obtain the \textit{ParInfo} for the current parameter \textit{Par} (line 8). We then fetch the current value of the parameter (line 9). If the parameter is non-modifiable, we randomly select a value from its predefined options in the \textit{Description} (lines 10-11). Otherwise, we proceed with further generation.

We then determine whether the type of \textit{Par} can be modified (line 13). If not, we retain the original type (line 14); if it has dependencies from previous parameters, we assign its type accordingly based on those dependencies (lines 15-16). Otherwise, we randomly select a type from the available options (lines 17-18).

A similar procedure applies to the size attribute: if the size of \textit{Par} can be modified (line 20), we keep the original size (line 21); if there are dependencies from previous parameters, we accordingly determine the size based on those dependencies (lines 22-23). If there are no dependencies, we select a size within the allowable range. (lines 24-25). 

Subsequently, we randomly choose one of the three predefined value strategies (line 27). Using this information, we generate a new parameter value \textit{New\_Arg} and update \textit{New\_Args} accordingly (lines 29-30). Once all parameters in \textit{Pars} are processed, we execute the API test with the generated input case \textit{New\_Args} (line 32). If a bug occurs, we log the test case and API for further analysis (lines 33-34), continuing the process until the generation limit is reached.

\subsection{Oracle}

\label{sec:bug}
In this section, we elaborate on our methodology for leveraging generated fuzzing outputs to test OpenCV library, employing both generic and OpenCV-specific oracles for bug detection. Our primary approach relies on reliability testing oracles, as outlined in existing research\cite{dejaegher2007ruggedness}. We execute programs on the CPU to capture all outputs, facilitating bug detection. We focus on identifying three critical types of bugs, each indicative of significant reliability concerns for software systems.

\textbf{Crashes}: Our detection efforts focus on bugs manifesting as unexpected crashes during the fuzzing phase. These include system disruptions such as aborts, segmentation faults, extensive memory leaks, and bugs flagged by internal assertion failures (INTERNAL\_ASSERT\_FAILED). Such crashes are alarming as they highlight immediate stability issues and expose potential security vulnerabilities that could be exploited maliciously.

\textbf{NaN Values}: Another focus of our bug detection strategy is identifying unexpected Not a Number(NaN) values during computations. NaN values are particularly concerning in critical systems, as they can lead to unpredictable and hazardous behavior~\cite{odena2019tensorfuzz}. NaN bugs typically arise from invalid mathematical operations or unsafe operations causing overflow or underflow conditions.

\textbf{Exceptions}: Our analysis extends to detecting anomalies during execution. These include arithmetic bugs, resource access discrepancies (e.g., correct inputs leading to incorrect exceptions or PermissionError), and logical bugs causing exceptions like index out-of-range or type incompatibility. Detecting such exceptions is crucial as they indicate bugs in program logic or resource handling and highlight areas to strengthen application robustness to prevent data corruption or unstable behavior.

Throughout the fuzzing regimen, should any of the aforementioned bugs be detected, we ensure that all relevant inputs and APIs implicated in these bugs are logged.

\section{Evaluation}
\subsection{Research Questions}

\begin{itemize}
  \item[\textbf{RQ1:}] How effective is \tool\ in detecting bugs?
  \item[\textbf{RQ2:}] How do the fuzzing configurations \tool\ affect its effectiveness?
  \item[\textbf{RQ3:}] How does \tool\ compare to existing constraint extraction approaches?
\end{itemize}

For RQ1, we investigate whether \tool\ is capable of detecting real bugs in the OpenCV-python library. For these bugs, we present the bugs detected by \tool\ and confirmed by OpenCV. Additionally, we analyze the necessity of testing poorly-documented APIs by examining the proportion of detected bugs originating from such APIs.
For RQ2, we focus on the impact of the number of generation times and the generation strategies of \tool\ on code coverage. Here, we checked the line coverage of OpenCV-python with different numbers of test case generations as well as the effects of different generation strategies on bug detection. 
For RQ3, we assess the effectiveness of the constraint extraction method used in \tool\ by comparing it with state-of-the-art constraint extraction methods and evaluating its performance through testing.

\subsection{Experimental Setup}

\begin{figure}[t!]
\centering
\small
\begin{tikzpicture}[font=\small]
\definecolor{color1}{RGB}{173, 216, 230}
\definecolor{color2}{RGB}{255, 192, 203}
\definecolor{color3}{RGB}{144, 238, 144}
\definecolor{color4}{RGB}{255, 165, 0}
\definecolor{color5}{RGB}{255, 215, 0}
\definecolor{color6}{RGB}{147, 112, 219}
\definecolor{color7}{RGB}{240, 100, 100}

\newcommand{\piechart}[7]{
    \pgfmathsetmacro{\totalnum}{#1+#2+#3+#4+#5+#6+#7}
    \pgfmathsetmacro{\cumnum}{0}
    \foreach \value/\color in {#1/color1, #2/color2, #3/color3, #4/color4, #5/color5, #6/color6, #7/color7}{
        \pgfmathsetmacro{\newcumnum}{\cumnum + \value/\totalnum*360}
        \draw[fill=\color] (0,0) -- (\cumnum:1.4cm) arc (\cumnum:\newcumnum:1.4cm) -- cycle;
        \global\let\cumnum=\newcumnum
    }
}

\piechart{248}{6}{16}{39}{32}{8}{330}

\begin{scope}[shift={(2.0cm,1.1cm)}] 
    \foreach \y/\color/\name in {
        0/color1/no documentation,
        1/color2/stereo and video,
        2/color3/file,
        3/color4/creating objects,
        4/color5/no outputs,
        5/color6/strong dependencies,
        6/color7/we covered
        }{
        \draw[fill=\color] (0,-\y*0.4) rectangle (0.15,-\y*0.4+0.15);
        \node[anchor=west] at (0.4,-\y*0.4+0.12) {\name};
    }
\end{scope}
\end{tikzpicture}
\caption{OpenCV-python APIs}
\label{figure-pie}
\end{figure}
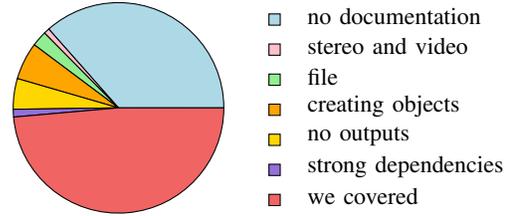

\noindent\textbf{Targeted library.} This work is targeted on the fuzzing of OpenCV-python (v4.9.0), which operates through an encapsulated Python interface that calls the underlying OpenCV C++ implementation.
In OpenCV-python (v4.9.0), there are a total of 679 APIs. As shown in Figure~\ref{figure-pie}, 248 of these are marked as "undocumented APIs," preventing the establishment of standardized API information using GPT-4. 6 APIs related to stereo and video are not supported by \tool. Additionally, 55 APIs used for reading files and 32 APIs without outputs are excluded from testing because their results cannot be directly measured. 10 APIs that are highly dependent on other APIs are also excluded since our method is focused on testing individual APIs. As a result, the remaining 330 APIs are selected for the experiments. Among these, 298 are well-documented, and the remaining 32 APIs are poorly-documented.

\label{selected_API}

\noindent\textbf{Testing budget.} For fuzz testing, \tool\ generates 600 input parameter lists for each API using standardized API information. The entire testing process takes a total of 3.2 hours.

\noindent\textbf{Environment.} We experiment on 64-core PC with 32GB RAM. 

\subsection{Metrics}

\noindent\textbf{Code Coverage.} Code coverage is a widely adopted test adequacy criterion in traditional software testing. Tests are unlikely to detect issues in portions of the code that they do not execute. Following the recent work on fuzzing of other Python libraries~\cite{gu2022muffin} and employ the \texttt{coverage.py} tool~\cite{coveragepy} to measure line coverage. However, since the underlying implementation of OpenCV-python is implemented in C++, using \texttt{coverage.py} does not track the coverage of the underlying C++ code. To address this, we set up an environment that enables us to collect both Python and C++ coverage data. Specifically, we employ \texttt{GCOV}~\cite{gcovmanual}, a coverage testing tool included with the GCC compiler, which allows us to measure the execution coverage of C++ code invoked via Python. As a result, our coverage analysis consists of two parts: (1) Python-level line coverage measured using \texttt{coverage.py}, and (2) C++-level line coverage of the underlying implementation, obtained using \texttt{GCOV}. This setup provides a comprehensive results of the tested portions of OpenCV-python, ensuring that both Python and native C++ code are counted.

\noindent\textbf{Detected bugs.} Following prior fuzzing research~\cite{xie2022docter}, we report the number of bugs identified during our testing process, providing insight into the comprehensiveness of our constraint extraction approach.

\noindent\textbf{The number of extracted constraints.} We count the number of constraints for each input parameter, providing insight into the comprehensiveness of our constraint extraction approach.

\noindent\textbf{The success rate of generation.} 
 We evaluate the effectiveness of our test case generation by executing the generated inputs on their corresponding APIs and measuring the percentage of test cases that run successfully.

\subsection{Baseline}

We evaluate the performance of the constraint extraction approach used in \tool\ by comparing it with state-of-the-art constraint extraction and testing methods on both poorly documented and well-documented APIs.
For poorly-documented APIs, we attempted to use DRONE~\cite{zhou2019drone}, a tool designed to detect and repair documentation deficiencies. However, DRONE is ineffective for APIs that only provide function signatures, as it does not handle cases where no detailed documentation exists.
For well-documented APIs, we evaluated DocTer~\cite{xie2022docter}, a closely related approach that extracts constraints from API documentation to fuzz deep learning libraries. However, DocTer is specifically designed for deep learning frameworks and cannot be directly applied to OpenCV. Despite this limitation, we successfully adapted DocTer’s constraint extraction component as a baseline for OpenCV documentation. After extracting constraints using DocTer, we applied our own constraint-processing mechanism to conduct testing.
In addition, we explored several fuzzing methods designed for deep learning libraries, including FreeFuzz~\cite{wei2022free}, TitanFuzz~\cite{deng2023large}, and FuzzGPT~\cite{deng2024large}. However, all of these methods face limitations when applied to the OpenCV library. FreeFuzz relies on historical open-source deep learning code for testing, TitanFuzz does not provide open-source code for generating test cases (only test cases for deep learning libraries), and FuzzGPT’s code is not open source, preventing its use for testing OpenCV. For Fuzz4ALL~\cite{xia2024fuzz4all}, we tried to apply it to OpenCV, but the generated test cases failed to meet OpenCV’s constraints, making it unsuitable for testing OpenCV library.

\section{Results and Analysis}

\subsection{RQ1: Effectiveness of \tool}

\subsubsection{Overall Results}

As shown in Table~\ref{tab:bugs1}, \tool\ detected 17 bugs in the OpenCV library across 330 APIs. These 17 bugs include 5 unexpected NaN value bugs, of which 1 has been confirmed, 3 crash bugs that caused the OpenCV to fail during testing, all of which have been confirmed, and 13 unexpected exception bugs, with 6 confirmed.
Among the 17 reported bugs, 10 were newly confirmed, 5 were previously known, and only 2 were false positives. Through our manual investigation, we found that the false positives were caused by errors in the documentation, which led to incorrect constraints being extracted. Of these 10 newly confirmed bugs, 5 have already been fixed, and the rest are being processed.

\begin{table}[t!]
\centering
\caption{Bugs Reported and Confirmed}
\label{tab:bugs1}
\resizebox{0.48\textwidth}{!}{
\begin{tabular}{lcccccc}
\hline
OpenCV & \textbf{Crash} & \textbf{NaN} & \textbf{Exception} & \textbf{Total} \\
\hline
\textbf{Reported} & 3 & 5 & 13 & 17 \\
\textbf{Confirmed} & 3 & 1 & 6 & 10 \\
\hline
\end{tabular}
}
\end{table}

\begin{figure}[t!]
\centering
\lstset{style=mystyle}
\begin{lstlisting}[language=Python, label = bug_nan]
point1 = np.float64( [[-2/2, -2/2], [-2, -2], [-2/2, 2/2], [-2, 2], [-2/2, 0]])
point2 = np.float64( [[0, -2/2], [-2/2, -2/2], [0, 0], [-2/2, 0], [0, -1/2]])
output = cv2.findHomography(point1, point2)
print(output)
\end{lstlisting}
\vspace{-3mm}
\lstset{style=mystyle_1}
\begin{lstlisting}[language=Python,  caption= Unexpected NaN, label=bug_nan_out]
output:
(array([[ inf,  nan,  inf],
       [ inf,  inf,  nan],
       [-inf,  nan,  nan]]), array([[1], [1], [1], [1], [1]], dtype=uint8))
\end{lstlisting}
\vspace{-5mm}
\end{figure}

The example of an unexpected NaN value in Listing~\ref{bug_nan_out} was generated by testing the input parameters of the API, as depicted. The bug resulted from \textit{Value Division}, that is, the random elements within the data are divided by an integer. We also manually verified that altering any single digit within this input data will not cause this bug. Regarding this bug, OpenCV has confirmed it has been fixed in the latest released version.

\begin{figure}[t!]
\centering
\lstset{style=mystyle}
\begin{lstlisting}[language=Python, label = bug_crash_2]
p1 = np.array([[[ 46.077175 , 228.66121  ]],
        ...
       [[243.1221   ,  60.95162  ]]], dtype=np.float32)
p2 = np.array([[[144.33624  , 247.15732  ]],
        ...
       [[ 39.08164  , 180.08517  ]]], dtype=np.float32)
out1, out2 = cv2.intersectConvexConvex(p1, p2, False)
\end{lstlisting}
\vspace{-3mm}
\lstset{style=mystyle_1}
\begin{lstlisting}[language=Python,  caption= Unexpected Crash, label=bug_crash_out_2]
output:
Process finished with exit code -1073740791 (0xC0000409)
\end{lstlisting}
\vspace{-5mm}
\end{figure}

The example of a crash bug is in Listing~\ref{bug_crash_out_2}. The execution did not yield any results but instead led to a crash, indicating an abnormal program termination. Such negative exit codes suggest that unexpected conditions forced the program to terminate prematurely, which is an issue of critical importance in software testing. OpenCV has confirmed that this bug has been fixed in the latest released version.

\begin{figure}[t!]
\lstset{style=mystyle}
\begin{lstlisting}[language=Python, label = bug_error]
P = np.array([[181.24588, ...], dtype=np.float32)
r = np.array([[0.9357548, ...], dtype=np.float32)
t = np.array([[69.32692 , ...], dtype=np.float32)
c = np.array([[214.0047, ...], dtype=np.float32)
d = np.zeros((3, 1), dtype=np.float32)
imagePoints, _ = cv2.projectPoints(P, r, t, c, d)
\end{lstlisting}
\vspace{-3mm}
\lstset{style=mystyle_1}
\begin{lstlisting}[language=Python,  caption= Exception, label=bug_error_2]
output:
cv2.error: OpenCV(4.9.0) D:\a\opencv-python\opencv-python\opencv\modules\calib3d\src\calibration.cpp:270: error: (-205:Formats of input arguments do not match) All the matrices must have the same data type in function cvRodrigues2.
\end{lstlisting}
\vspace{-5mm}
\end{figure}

Listing~\ref{bug_error_2} illustrates a bug encountered during the invocation of \texttt{\small cv2.projectPoints}, where an unexpected exception occurred despite our generated test cases fully adhering to the specified requirements. OpenCV has confirmed that this bug has been fixed in the latest released version.

The bugs presented above have been fixed by OpenCV developers, demonstrating the effectiveness of our generation strategy in monitoring and testing increasingly complex computer vision applications, ensuring that they operate in accordance with predefined safety considerations. The successful identification and resolution of these errors underscore the importance of employing a diversified generation strategy. This approach not only improves the adaptability of the system to abnormal inputs and edge cases, but also ensures reliability in extreme scenarios.

\subsubsection{Bugs in Poorly-Documented APIs}

We used GPT-4 to generate standardized API information for poorly-documented APIs, specifically testing 32 poorly-documented ones. Our testing identified 5 bugs from the poorly-documented APIs, with 2 confirmed. 
Although they made up only 9.7\% of the total APIs tested, poorly-documented APIs were accounted for 29.4\% of all detected bugs and 20\% of the confirmed ones. 
This indicates that APIs with insufficient documentation are more prone to bugs and less reliable. These findings highlight the need to pay particular attention to these APIs.

\begin{tcolorbox}[colback=gray!20, colframe=gray!60, sharp corners]
\small
\textbf{
\textit{
\tool\ detected 17 bugs in OpenCV, of which 10 were confirmed and 5 have been fixed. Despite constituting only 9.7\% of the tested APIs, poorly-documented ones accounted for 29.4\% of the detected bugs, highlighting their higher defect rate and \tool's effectiveness in identifying such bugs.
}
}
\end{tcolorbox}

\subsection{RQ2: Ablation Study}
\label{RQ2:Ablation study}

\begin{figure}[t!]
    \centering
    \includegraphics[width=1\columnwidth]{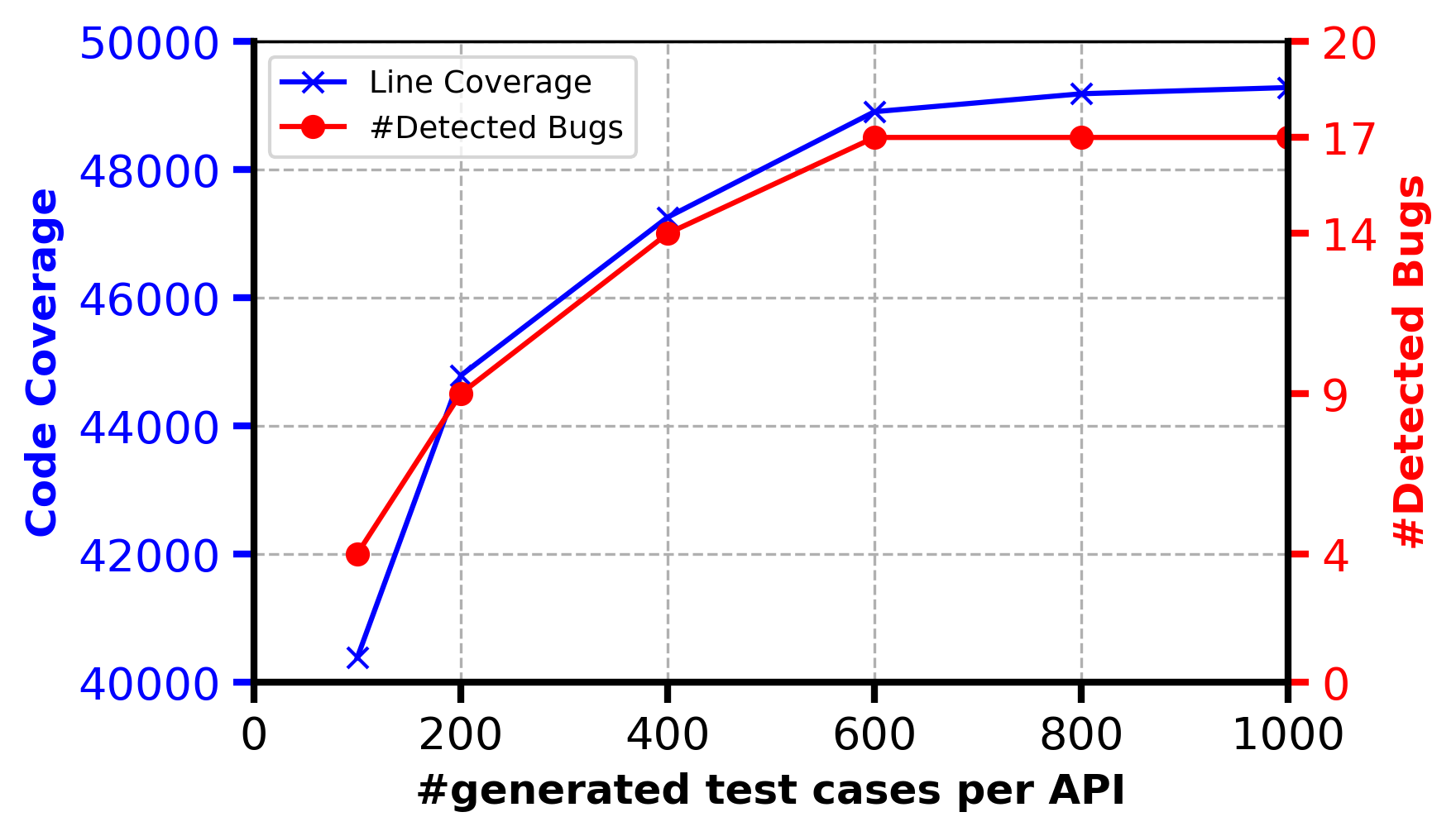}
    \caption{
        Code Coverage and Bug Detection Trend for OpenCV.
    }
    \label{fig:coverage_and_bugs}
\end{figure}

\subsubsection{Code Coverage and Bug Detection}
\label{RQ2:code_coverage}
{We demonstrate the effectiveness of \tool\ in improving code coverage and detecting bugs when generating varying numbers of test cases per API.  Figure~\ref{fig:coverage_and_bugs} presents the results: the x-axis represents the number of test cases generated per API (ranging from 100 to 1000), and the left y-axis denotes the total code coverage achieved across all tested APIs (i.e., the union of all coverage sets), and the right y-axis indicates the number of detected bugs.
Note that the starting point represents the code coverage achieved through the direct execution of the original test inputs without any modification. The results show that increasing the number of generated test cases leads to improved code coverage, which validates the effectiveness of our generation strategy. However, coverage gains plateau around 600 test cases per API, suggesting this number as a cost-effective threshold.
Similarly, bug detection also saturates: increasing the number of generated test cases per API improves bug-finding capability only up to a certain threshold. Specifically, 200 and 400 test cases revealed 9 and 14 detected bugs, respectively, while 600 test cases uncovered all 17 bugs identified in our evaluation. Beyond 600 test cases, no further bugs were discovered, while computational overhead continued to rise. }

\begin{figure}[t!]
    \centering
    \includegraphics[width=1\columnwidth]{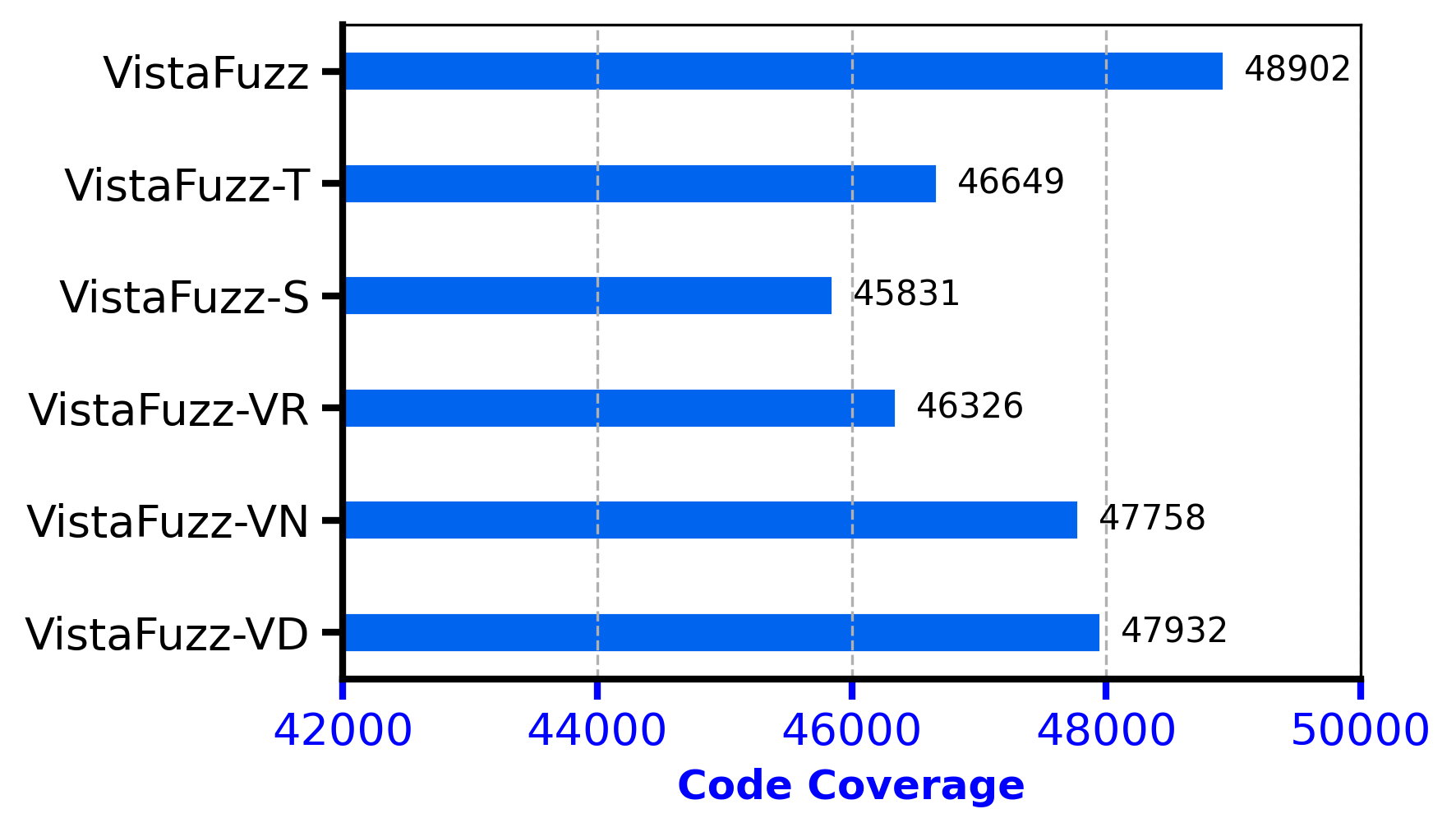}
    \caption{
        Code Coverage of Different Generation Strategies.
    }
    \label{fig:coveragecompare}
\end{figure}

\subsubsection{Generation Strategies}
\label{RQ2:bug detection}
Next, we analyze the impact of different generation strategies. As mentioned in Sec~\ref{sec:gs}, we designed three strategies to modify input parameters—\textit{Type}, \textit{Size}, and \textit{Value}—to enhance bug detection. The \textit{Value} strategy consists of three sub-strategies: \textit{Adding Noise}, \textit{Random Masking}, and \textit{Division}.

\begin{figure}[t!]
  \centering
  \includegraphics[width=0.61\linewidth]{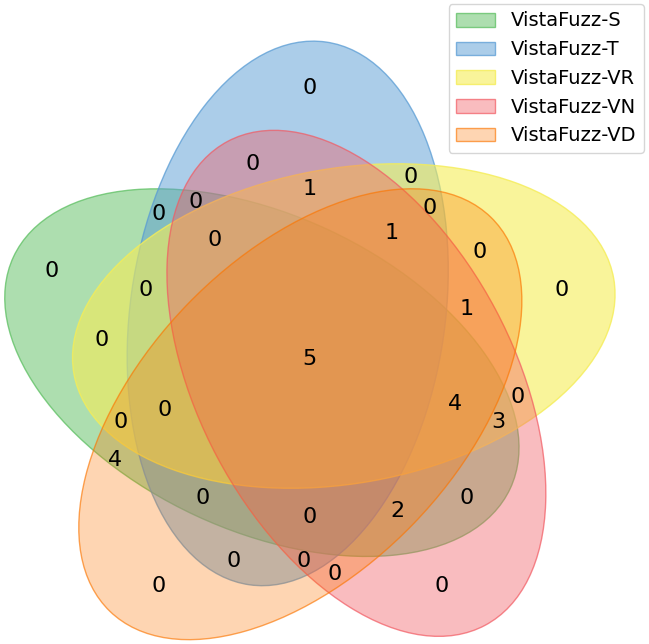}
  \caption{Bug Detection by Different Generation Strategies.}
  \label{figure:venn}
\end{figure}

For each strategy, we ensured that the number of generations was consistent.
Here, we investigate further the impact of each generation strategy. To this end, we have five variants of \tool, namely \tool-T (with \textit{Type} disabled), \tool-S (with \textit{Size} disabled), \tool-VR (with \textit{Random Masking} disabled), \tool-VN (with \textit{Adding Noise} disabled), and \tool-VD (with \textit{Division} disabled). From the Figure~\ref{fig:coveragecompare}, we can make the following observations. Under a limit of generating 600 test cases, first, the complete version of \tool\ outperforms all other variants studied in terms of code lines covered, which highlights the importance and necessity of all generation strategies implemented in \tool. Secondly, as for the bugs detected, the results are as shown in Figure~\ref{figure:venn}. Clearly, for OpenCV-python, disabling any strategy results in missing certain bugs, thus proving the necessity of our generation strategies.

\begin{tcolorbox}[colback=gray!20, colframe=gray!60, sharp corners]
\small
\textbf{
\textit{
\tool\ achieves cost-effective code coverage by generating 600 test cases per API. Additionally, it outperforms all ablated variants, confirming that all generation strategies are necessary for bug detection and coverage.
}
}
\end{tcolorbox}

\subsection{RQ3: Comparison with Existing Approaches}
Our method for extracting constraints is based on the intermediate representation of standardized API information parsed by GPT-4, from which we further extract constraints between different parameters of each API. 
To facilitate comparison with existing approaches, we have established two key metrics: one is the number of constraints extracted for APIs, and the other is the success rate of the test cases generated based on these constraints. For the first metric, we count the number of constraints for each input parameter under each API. For the success rate of the test cases, we apply these input cases to the corresponding API and check whether the API will throw an error due to incorrect input parameters. Through this approach, we can effectively assess the suitability and stability of both the constraint extraction method and the generated test cases.

\begin{table}[t!]
\centering
\caption{Comparison with DocTer}
\label{tab:comparison}
\resizebox{0.48\textwidth}{!}{
\begin{tabular}{lccccc}
\hline
  & \textbf{\# Cov} & \textbf{\# Bug} & \textbf{\# EC} & \textbf{\% SRG} \\ \midrule
\textbf{\tool} & 48,902 & 17 & 2,797 & 99.39\% \\ 
\textbf{DocTer} & 7,829 & 3 & 528 & 8.74\% \\ 
\bottomrule
\end{tabular}
}
\end{table}

Since DocTer cannot be applied to poorly-documented APIs, we applied DocTer on well-documented APIs and constraint extraction component to parse the documentation into dependency trees, from which constraints were then extracted. 
These constraints are then used for fuzzing the target APIs, the same as the second phase of \tool. As code coverage~(\#Cov), detected bugs~(\#Bug), the number of extracted constraints~(\#EC), and the success rate of generation~(\#SRG) shown in Table~\ref{tab:comparison}, this process extracted 528 constraints, while \tool\ extracted 2,797 constraints. Additionally, due to DocTer having no consideration of dependencies between different input parameters, the valid test case generation rate was only 8.74\%, whereas \tool\ achieved 99.39\%. In particular, the remaining failed test cases were due to discrepancies between the document description and the code implementation.
In the testing of these generated test cases, only 3 bugs were discovered, and all these bugs have already been covered by our method.
Ultimately, our method \tool\ reached a line coverage of 48,902, compared to only 7,829 lines covered by fuzzing based on DocTer's constraint extraction component.

\begin{tcolorbox}[colback=gray!20, colframe=gray!60, sharp corners]
\small
\textbf{
\textit{
\tool\ extracts 2,797 constraints compared to DocTer’s 528, achieves a 99.394\% valid test case rate, and delivers six times higher code coverage. It detects all bugs found by DocTer and additionally extends to poorly-documented APIs.
}
}
\end{tcolorbox}

\section{Discussion}

Several components of \tool\ are specifically designed for testing OpenCV, including extracting constraints and dependencies from standardized API information to generate valid input parameter values, and the formulation of generation strategies that meet OpenCV’s API requirements. However, the importance of our idea transcends the scope of OpenCV, which can also extend to testing across libraries in various dynamic-type languages. This broader applicability is anchored in the methodology of utilizing library documentation to inform the fuzzing process.

\textbf{Extensibility}.
\tool's strategy of generating standardized API information from documentation is language- and library-agnostic, making it easily extensible to other Python projects and libraries in dynamically typed languages. Its key steps, documentation parsing, constraint extraction, and dependency modeling, can be extended to other libraries by updating input validation rules and type converters to align with the data types and conventions of the target library. 

\textbf{Choice of LLM}.
We selected GPT-4 due to its demonstrated strength in code comprehension and natural language tasks, as supported by recent research~\cite{chen2021evaluating}. In early pilot experiments, alternative open models (such as GPT-3.5, Llama-2) exhibited noticeably lower accuracy and required more manual correction, especially when parsing complex or ambiguous documentation.

{\textbf{Threats to Validity}}.
The primary threats to validity stem from the implementation of \tool. To mitigate this, we performed extensive tests and code reviews to confirm its correct implementation. Additionally, our approach relies on documents as the primary data source, which may introduce bias due to potential limitations in document selection, representativeness, or completeness. To address this, we conducted a rigorous examination to ensure data integrity and usability. Furthermore, our evaluation focuses solely on OpenCV. While OpenCV is widely used, its unique parameter formats and API design may limit the generalizability of \tool\ to other libraries without further adaptation.

\textbf{Future Work}.
Currently, as a document-guided fuzzing technique, \tool\ does not infer constraints for APIs that have no documentation. A promising direction for future work is to incorporate source code analysis in such cases. For undocumented APIs, relevant information such as function signatures and type hints can be extracted directly from the source code. By combining documentation analysis with source-code-based inference, it may be possible to infer constraints even for undocumented APIs. This hybrid integration presents a promising avenue for improving the generalization and applicability of the approach.

\section{Related Work}

Fuzzing~\cite{miller1990empirical} is an automated testing technique that executes the target system with random or invalid inputs to uncover anomalies such as crashes and hangs. It has been widely used in various domains, including operating systems~\cite{chen2022sfuzz}, network protocols~\cite{pham2020aflnet}, web applications~\cite{atlidakis2019restler}, and APIs~\cite{deng2022fuzzing}. 
Recent advancements in LLMs have enabled their use in test case generation. For example, TitanFuzz~\cite{deng2023large} modifies API inputs and outputs in deep learning libraries to uncover bugs, while FuzzGPT~\cite{deng2024large} generates edge cases based on historically bug-inducing code. Fuzz4ALL~\cite{xia2024fuzz4all} applies LLMs to generate test cases for compilers and virtual machines, relying on the LLM’s understanding of the system, and CHATAFL~\cite{meng2024large} focuses on protocol fuzzing by generating message sequences to explore different protocol states. 
{LISP~\cite{li2024llm} applies LLMs to input space partitioning for library APIs, aiming to maximize coverage by dividing input domains.}
However, these methods heavily depend on the LLM’s prior knowledge of the target library, making it challenging to generate effective test cases for libraries that the LLM lacks sufficient understanding of.

For libraries or systems where LLMs cannot directly generate effective test cases, document-guided testing provides a reliable approach by extracting constraints from software documentation to generate more structured test cases. Traditional approaches have utilized documentation~\cite{lv2020rtfm} and annotations~\cite{zhou2018automatic} to identify inconsistencies between specifications and implementations. Some methods transform specifications into assertions~\cite{liu2014automatic} and oracles~\cite{motwani2019automatically}, while others rely on manually crafted rule-based extraction~\cite{blasi2018translating}. DocTer~\cite{xie2022docter} applies sub-tree mining and associative rule learning to extract API constraints, but existing methods often overlook dependencies between parameters and assume complete documentation. Our approach addresses these limitations by using LLMs to parse official documentation into a standardized format and to extract parameter constraints and dependencies, enabling more effective test case generation, even for poorly documented APIs.

{Unlike most deep learning APIs, which are typically designed to operate on a single main tensor with minimal cross-parameter dependencies, OpenCV APIs often require multiple parameters to satisfy interdependent constraints, such as image size, type consistency, or region bounds. This presents a unique challenge for automated test input generation. As shown in our evaluation (Table~\ref{tab:comparison}), methods like DocTer~\cite{xie2022docter}, which are effective on testing deep learning libraries, struggle with OpenCV library due to their limited modeling of parameter dependencies. These findings underscore the need for approaches tailored for vision libraries, where parameter interactions are critical.
}

\section{Conclusion}

This paper introduced \tool, a pioneering approach that utilizes LLMs to enhance fuzzing in the OpenCV library by learning from well-documented APIs and improving the handling of poorly-documented ones. By generating standardized API information and extracting constraints and dependencies to generate effective test inputs, \tool\ successfully detected 17 bugs in 330 APIs, where 10 bugs have been confirmed and 5 of them have been fixed. 


\balance
\bibliographystyle{IEEEtran}
\bibliography{main}

\end{document}